# Origin of resistivity minima at low temperature in ferromagnetic metallic manganites

P.R. Sagdeo*, R.J. Choudhary and D.M. Phase

UGC-DAE Consortium for Scientific Research, University Campus, Khandwa Road, Indore-452001, INDIA.

**Abstract**

The resistivity and magneto-resistance measurements were carried out on thin film of $La_{0.7}Ca_{0.3}MnO_3$ to investigate the possible origin of low temperature resistivity minimum observed in these samples. We observed large hysteresis in the magnetoresistance at low temperature (5K) and the sample current 'I' has large effect on resistivity minima temperature. The observation of hysteresis at low temperatures suggests the presence of in-homogeneity at low temperatures. These in-homogeneities consist of regions of different resistive phases. It appears that the high resistive phase prevents the tunneling of charge carriers between two low resistive regions and thus giving rise to the resistivity minimum in these samples.

* Email: sagdeo@csr.ernet.in & prsagdeo@yahoo.com

## Introduction

The electrical transport in perovskite manganite is one of the hottest topics in condensed matter physics. The phenomenon of colossal magnetoresistance (CMR) [1], spin polarized transport [2], scattering of electrons from grain boundaries [3], low temperature resistivity minima [4], low temperature magneto-resistance/tunneling through the grain boundaries [5], nature of charge carriers in various regimes [6], the interlink between the electrical properties and magnetic state of the sample [7] etc, are still not well understood through a common frame work. Extensive theoretical efforts have been made right from 1951 [8-15], but the electrical transport in manganites still remains the challenging problems of this field.

The ferromagnetic metallic manganites show the minima in the resistivity at low temperatures [4, 16-20]. This minimum in the resistivity has been attributed to inter grain anti-ferromagnetic coupling [16], enhanced e-e interactions [17], Kondo like scattering [18,19] and also to the quantum interference effects including weak localization [20]. Thus, in the available literatures, varieties of models are available to explain the phenomenon of low temperature resistivity minimum.

In this brief report we present the resistivity and magneto-resistance studies on the oriented thin film of $La_{0.7}Ca_{0.3}MnO_3$. Our result suggests that the minimum in the resistivity is due to the different resistive phases present in these samples at low temperatures. It appears that the high resistive phase prevents the tunneling of charge carriers between two low resistive regions and thus, giving rise to the resistivity minimum in these samples. The volume fraction of these different resistive states can be tuned by application of magnetic field and also by changing the sample current 'I'.

**Experimental**

The polycrystalline bulk target of $La_{0.7}Ca_{0.3}MnO_3$ (LCMO) has been prepared by standard solid-state reaction route [23]. This target is used to grow the oriented thin film of $La_{0.7}Ca_{0.3}MnO_3$ on single crystal $LaAlO_3$ substrate, using Pulsed Laser Deposition (PLD) technique. During the deposition; oxygen partial pressure was kept at 200m-torr, target to substrate distance was kept at 4 cm and substrate temperature was maintained at $600^0$C. The prepared sample is characterized using powder x-ray diffraction and scanning electron microscopy. The temperature dependence of DC resistivity was carried out using Van Der Pauw four-probe resistivity measurement technique [3,24]. The magnetoresistance measurements on these samples were carried out in the presence of magnetic filed from 0 to 5 Tesla.

**Results and Discussions**

The structural properties of the as grown samples were studied using x-ray diffraction. This study suggests the highly oriented nature of the grown sample. The surface morphology of the prepared sample is studied using JEOL-5600 Scanning Electron Microscope (SEM). The SEM studies suggest the uniform growth of the prepared sample. This well characterized sample was used for resistivity and magnetoresistance measurements.

Figure1 shows the temperature dependence of resistivity for the studied samples. The sample shows insulator to metal transition ($T_{IM}$) at ~250K and a minima at low temperatures (~13 K as shown in the inset). In order to see the effect of magnetic field on the resistivity minima, we have carried out the temperature dependence of resistivity measurements in the presence of magnetic fields ranging from 0 to 5 Tesla. During the measurements we have observed that the sample current ‘I’ has large effect on the resistivity as well as on resistivity minima temperature ($T_{Rmin}$). Therefore, keeping in view the non-linear I-V characteristics of manganites [25], sample current (I) was kept constant during all the measurements. Figure 2 shows the temperature dependence of resistivity in the presence of various magnetic fields. In order to check the effect of magnetic field on the resistivity and $T_{Rmin}$ we have plotted the normalized resistivity (with respect to $\rho_{min}$ value) and the variation of $T_{Rmin}$ as a function of magnetic field, shown in the insets of Figure 2. From inset (a) of Figure 2 it is clear that resistivity enhances due to the application of magnetic field below resistivity minima and $T_{Rmin}$ increases with increasing magnetic field (see inset (b) Figure 2). It is important to note that these results are exactly opposite to the results reported by Xu *et al.* [19] where $T_{Rmin}$ decreases with magnetic field. The results reported by Xu *et al.* support the presence of Kondo like scattering in their samples. From Figure 2 it is clear that the application of magnetic field decreases the resistivity of the sample, therefore it may be possible that one may increase the sample current to keep noise level unchanged. The effect of sample current on the resistivity and $T_{min}$ is demonstrated in Figure 3. It shows the data for 0 Tesla and 5 Tesla for different value of sample currents. If one compare the resistivity for 0 Tesla (I=0.01mA) and 5 Tesla (0.1mA) then these results are comparable with those reported by *Xu et al.* our findings suggests that the sample current have large effect on resistivity minima, further our experiment confirms that the LCMO have non-ohmic I-V in this temperature range. Therefore, in order to get rid of sample current on the resistivity and $T_{Rmin}$ we have kept the sample current constant throughout the measurements and analyzed the data with the help of models available in the earlier reported literatures [16-20]. We have analyzed the resistivity data in this temperature range, considering *e-e* interaction, as have been suggested in the literature [17]. It was found that even though the resistivity data can be well fitted, considering *e-e* interaction but the coefficient of *e-e* interaction term is too large to account such interactions [26].

In the case of manganites it is known that the electronic [27], magnetic [28] and structural [29] inhomogeneity plays crucial role on the transport and the presence of such in-homogeneities is reflected in the hysteresis measurements. Therefore, in order to further investigate the origin of resistivity minima observed in these materials, we have studied the hysteresis in the resistivity & magneto-resistance below and above resistivity minima temperature.

Figure 4 shows the variation of magneto-resistance as a function of applied magnetic fields at 5K and 80K with positive and negative cycles. From the figure-4 it is clear that the sample shows very large hysteresis in the magneto-resistance below resistivity minima temperature (5K), whereas, it does not show the considerable hysteresis above resistivity minima temperature (80K). We have analyzed these resistivity and magneto-resistance measurements in the framework of 'quantum interference effects-with spin orbit interactions' [20], but the observation of large hysteresis in the magneto-resistance could not be understood in this framework [30]. The presence of hysteresis in the magnetoresistance data (5K) suggests the presence of in-homogeneity in the sample in this temperature range. Recently, Wagenknecht *et al.* using laser microscopy have shown the presence of different resistive state in this temperature range [31] and they have attributed the presence of different resistive state to the motion and flipping of magnetic domain wall. These authors have also reported large hysteresis in the domain-wall/grain-boundary resistance as function of magnetic field. Thus, the appearance of large hysteresis below the resistivity minima temperature might be due to the different resistive state present in the sample. It appears that the high resistive phase prevents the tunneling of charge carriers between two low resistive regions and thus, giving rise to the resistivity minimum in these samples. Thus, the appearance resistivity minima observed in these samples may be due to the presence of different resistive phases at low temperatures, and may not be due to the interactions effect as suggested in the earlier literatures [16-20].

In conclusion, we have very carefully carried out the resistivity and magnetoresistance studies on the thin film of $La_{0.7}Ca_{0.3}MnO_3$, occurrence of large hysteresis in the magnetoresistance measurements below resistivity minima temperature (5K)) suggests the presence of in-homogeneity in the sample in this temperature range. These in-homogeneities consist of regions of different resistive phases. The high resistive phase prevents the tunneling of charge carriers between two low resistive regions and thus giving rise to the resistivity minimum in these samples. We further observe that the current have large effect on the resistivity minima temperature. Thus, the present experiment confirms the role of complex magnetic and orbital interactions on the transport properties of these oxides.

**Acknowledgement:**

Authors sincerely thank Dr. Rajeev Rawat for valuable discussions and resistivity measurements. Authors are also thankful to Dr. P. Chaddah and Prof. Ajay Gupta for their encouragement and interest in the work. One of the authors PRS thank DST Government of India for financial support.

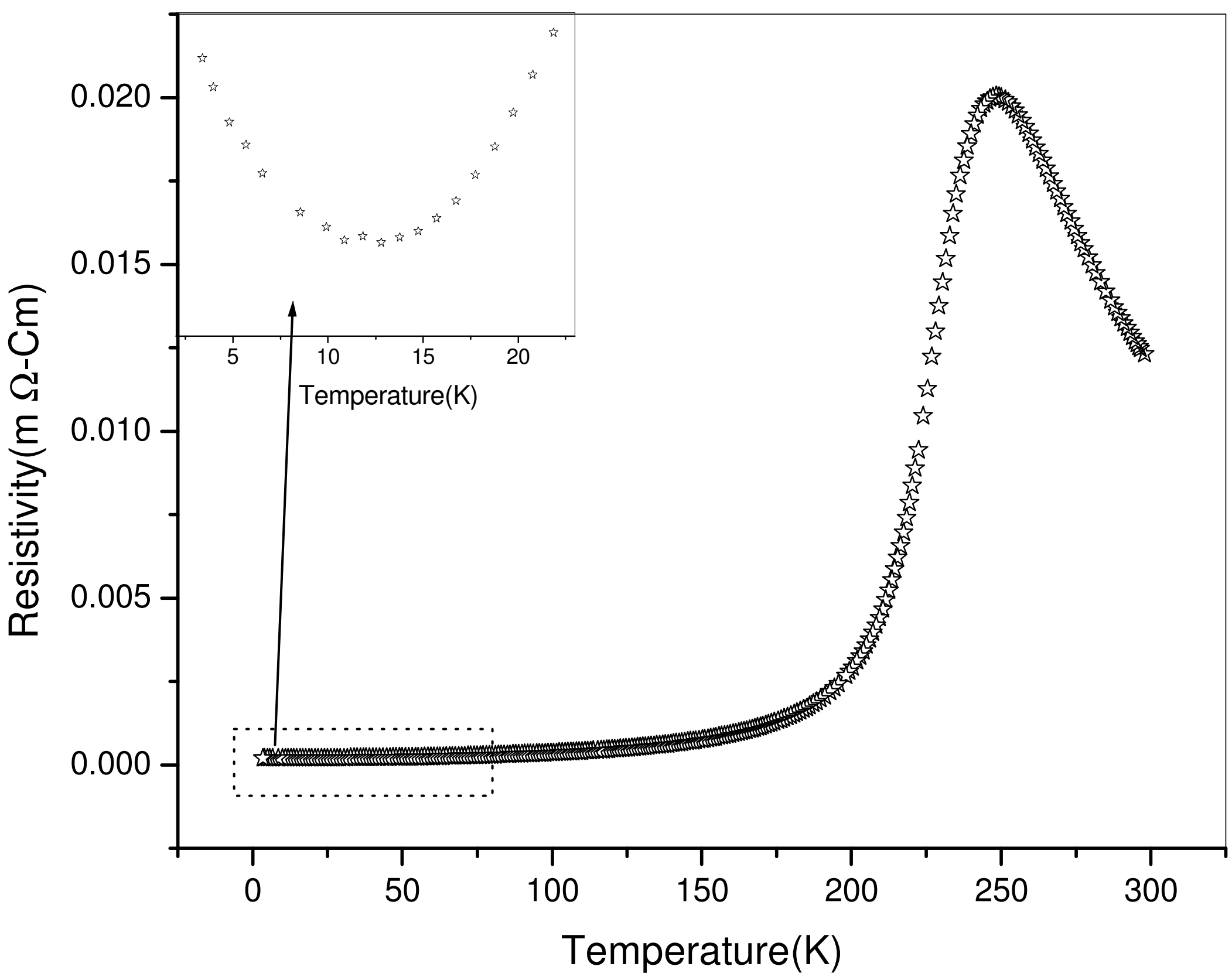

Figure 1

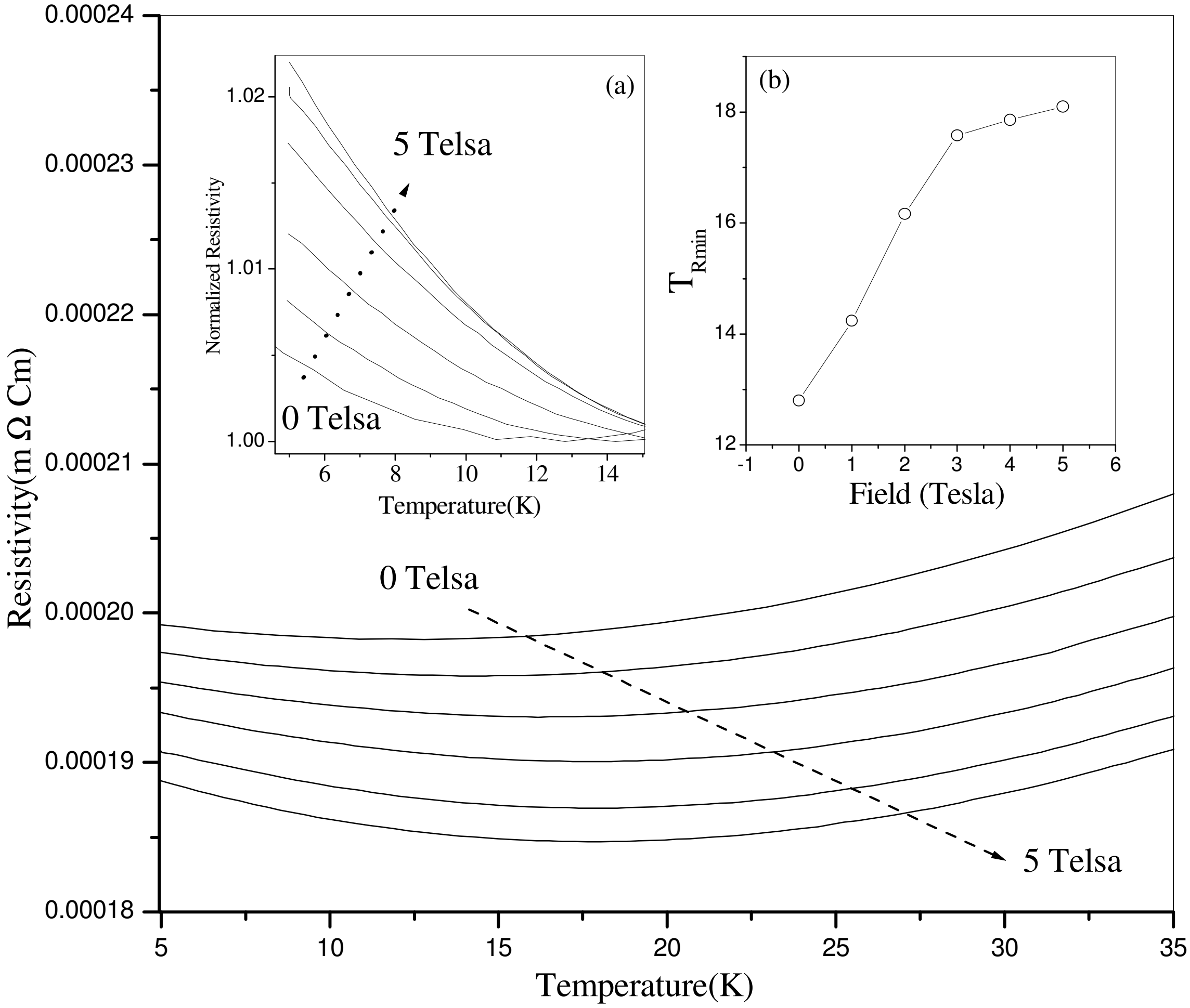

Figure-2

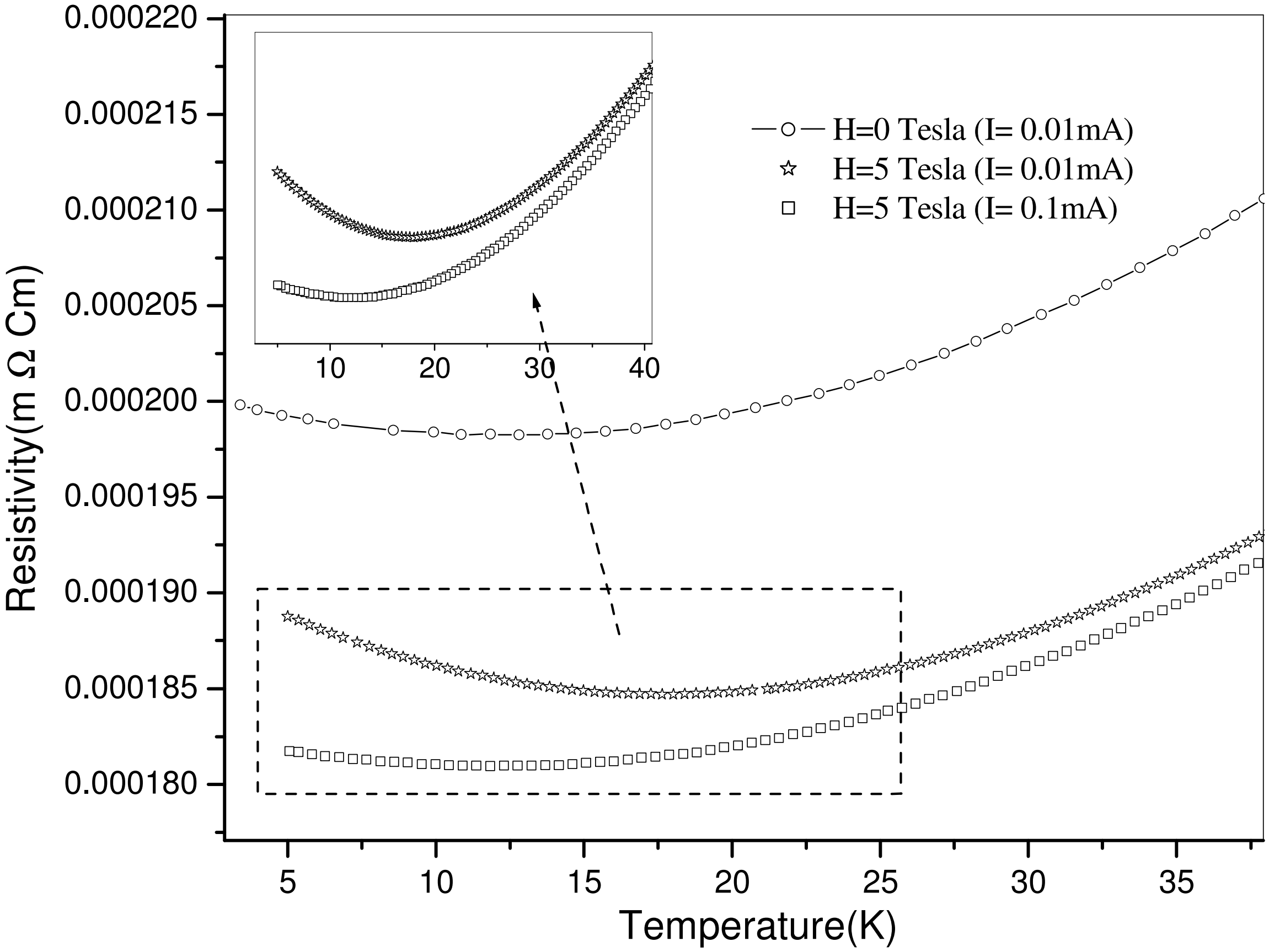

Figure-3

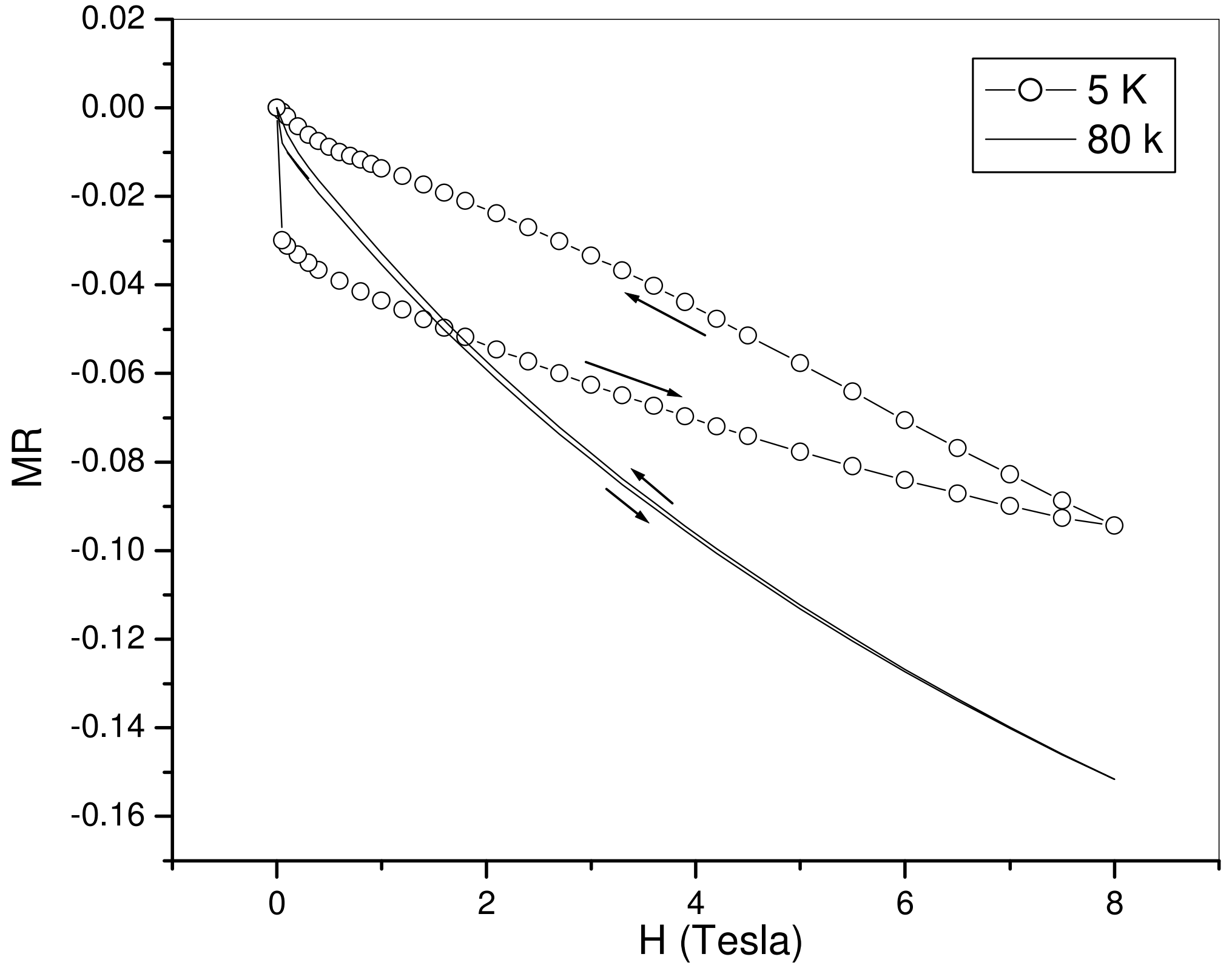

Figure-4

Figure captions:

Figure-1: The temperature dependence of resistivity for; $La_{0.7}Ca_{0.3}MnO_3$ thin film sample showing, insulator to metal transition at ~250K and the inset show the occurrence of minima at low temperatures ~13 K.

Figure-2: Variation of temperature dependence of resistivity as a function of applied magnetic field, inset (a) is the normalized resistivity with respect to $\rho_{min}$ value and (b) shows the variation of $T_{Rmin}$ as a function of magnetic field.

Figure-3: The effect of sample current 'I' on the resistivity and $T_{Rmin}$ for $La_{0.7}Ca_{0.3}MnO_3$ thin film sample.

Figure-4: Hysteresis in magneto-resistance as a function of applied magnetic fields at 5K and 80K with positive and negative cycles.